\documentclass[pre,aps,floatfix,showpacs,preprint,preprintnumbers,superscriptaddress]{revtex4-1}
\usepackage{graphicx}
\usepackage{dcolumn}
\usepackage{bm}
\usepackage{amssymb}
\usepackage{pifont}
\usepackage{amsmath}
\usepackage{times}
\usepackage{color}

\newcommand{\ii}{\mathrm{i}}
\newcommand{\dd}{\mathrm{d}}
\newcommand{\w}{\omega}
\newcommand{\W}{\Omega}

\begin{document}

\title{On the equivalence of phase-oscillator and integrate-and-fire models}

\author{Antonio Politi}
\email{a.politi@abdn.ac.uk}
\affiliation{Institute for Complex Systems and Mathematical Biology and SUPA,
University of Aberdeen, Aberdeen AB24 3UE, United Kingdom}

\author{Michael Rosenblum}
\email{mros@uni-potsdam.de}
\affiliation{Department of Physics and Astronomy, University of Potsdam,  
Karl-Libknecht-Strasse 24/25, 14476 Potsdam, Germany}

\date{\today}
\begin{abstract}
A quantitative comparison of various classes of oscillators (integrate-and-fire, Winfree, 
and Kuramoto-Daido type) is performed in the weak-coupling limit for a fully  connected 
network of identical units. An almost perfect agreement is found, with only tiny 
differences among the models. 
We also show that the regime of self-consistent partial synchronization [SCPS]
is rather general and can be observed for arbitrarily small coupling strength in any model class. 
As a by-product of our study, we are able to show that an integrate-and-fire model with a generic
pulse shape can be always transformed into a similar model with $\delta$-pulses 
and a suitable phase response curve.
\end{abstract}


\maketitle

\section{Introduction} 
Many studies of neural networks and, generally, of coupled oscillators are based 
on the assumption that the relevant dynamical properties can be reproduced 
by restricting the study to dynamical systems characterized by a single variable:
the {\it phase}. In spite of its simplicity, this setup is indeed able to produce
a wealth of nontrivial phenomena, ranging from the synchronization
transition \cite{Kuramoto-75,Kuramoto-84,Acebron-etal-05}, to 
self-consistent partial-synchronization \cite{Kuramoto-91,vanVreeswijk-96,Mohanty-Politi-06},
and including chimera states \cite{Kuramoto-Battogtokh-02,Abrams-Strogatz-04}, 
to name just a few.

The first such model was proposed by Winfree in 1967 to characterize
biological rhythms \cite{Winfree-67,Winfree-80}. In the weak-coupling limit, it may reduce to the
famous Kuramoto model \cite{Kuramoto-75,Kuramoto-84,Sakaguchi-Kuramoto-86},
that is currently much used to investigate the synchronization properties of various setups.
While in the Winfree model the coupling depends on the absolute value of the oscillator phases,
in the Kuramoto model it depends sinusoidally on phase-differences. In fact, the Kuramoto
model has been generalized to the so-called Kuramoto-Daido 
model~\cite{Daido-92a,*Daido-93a,*Daido-96}, 
where the coupling is a generic function of the phase difference.

Independently, yet another class of oscillators is being investigated: 
the so-called pulse-coupled integrate-and-fire oscillators.
Here, a single phase-like variable, describing the membrane potential, 
increases linearly until it reaches a threshold, is thereby reset to
some specific value, and simultaneously triggers the emission of a pulse that
is responsible for the mutual coupling. The effect of the pulse onto
the receiving oscillator is quantified by the phase response curve.
The simplest of such models was proposed in the context of heart activity~\cite{Peskin-75}, 
but is nowadays quite popular in computational neuroscience, where it is widely used 
to clarify the collective dynamics of neural circuits~\cite{Mirollo-Strogatz-90}. 
A similar and much used model is the leaky integrate-and-fire (LIF) neuron,
introduced by L. Lapicque in 1907, even before physiological mechanisms 
of pulse transmission were understood \cite{Lapicque-07}. There, the membrane
potential evolves exponentially rather than linearly in time.

Nowadays, whenever oscillatory phenomena have to be investigated, 
integrate-and-fire and Kuramoto-like
models are the most used setups, but it is not clear to what extent the resulting
phenomenology is typical of the selected model. 
A prominent example to illustrate the lack of a general framework is
self-consistent partial synchronization (SCPS), a regime where identical oscillators are neither 
locked, nor completely asynchronous.
Kuramoto~\cite{Kuramoto-91} found evidence of SCPS in a network of identical 
LIF oscillators in the
presence of noise and delayed $\delta$-pulses.
Later, van Vreeswijk observed and analysed this regime in an ensemble of LIF 
oscillators coupled through
smooth pulses and in the absence of external noise~\cite{vanVreeswijk-96}.
SCPS may also arise in the simple Kuramoto-Sakaguchi model~\cite{Sakaguchi-Kuramoto-86} 
(sine coupling with a phase-shift) 
but only for a particular value of the phase-shift, when it is marginally stable.
The onset of a robust SCPS regime is, however, possible in a Kuramoto-Sakaguchi-like setup, 
under the condition that the phase-shift parameter of the sine function depends on the 
order parameter and the coupling 
strength~\cite{Rosenblum-Pikovsky-07,*Pikovsky-Rosenblum-09}. This model can be obtained 
as a phase approximation of nonlinearly coupled Stuart-Landau oscillators.  

Another example of differences among the various setups is emergence of 
the irregular collective
dynamics in an ensemble of heterogeneous LIFs with delayed 
$\delta$-pulses~\cite{Luccioli-Politi-10}.
The setup is superficially analogous to the Kuramoto ensemble, but chaotic collective 
oscillations are not possible in the latter 
model~\cite{Watanabe-Strogatz-93,*Watanabe-Strogatz-94,Pikovsky-Rosenblum-11}.

In this paper we compare the various model classes in the minimal setup of identical 
globally coupled oscillators. In order to carry on a meaningful quantitative
analysis, three models (A, B, and C) are selected as follows.
Model A is the ensemble of LIF neurons extensively studied in Ref.~\cite{Abbott-vanVreeswijk-93}.
By then following~\cite{Golomb-Hansel-Mato-01}, model A is mapped, in the weak-coupling limit,
onto a Winfree-type ensemble of oscillators, yielding model B. 
Finally, model C is obtained as an approximate reduction of model B to a
Kuramoto-Daido ensemble.

Our studies reveal that the scenario emerging from the three models is substantially equivalent
with a couple of quantitative discrepancies which concern the fully synchronous regime: 
(i) the dependence of the period on the coupling strength is different in model A already
at the leading order; (ii) its stability differs in model C.
Finally, the equivalence between models A,B, and C implies that a generic LIF model 
with pulses of finite width can be mapped onto a model of pulse-coupled
oscillators and $\delta$-like pulses which can be more easily simulated with 
event-driven algorithms. To test this conjecture a model of the latter type
is introduced (model D).

More specifically, in section II, we introduce the various model classes, 
discuss their mutual relationships,
and briefly recall the most common asymptotic regimes. Section III is devoted to a
quantitative comparison of the models A, B, C, and D: 
in practice the analytically estimated stability spectra of the splay and 
synchronous states, as well as the numerically obtained features of the SCPS 
are mutually compared. 
Section IV is devoted to a perturbative analysis of SCPS in the Kuramoto-Daido setup. 
The resulting frequency of SCPS are found to be in excellent agreement with the numerical findings.
The main results and the open problems are summarized in section V. 
Finally, the many technical details related to the stability analysis
of the different regimes in the various models are confined to five appendices.

\section{Dynamical regimes and model classes}

As it is well-known, globally coupled ensembles of identical oscillators can exhibit two 
highly symmetric regimes: (i) a fully synchronized state, where all the oscillators are
characterized by the same phase at any time and (ii) an asynchronous regime, 
also called {\it splay state}, 
where the phases are uniformly distributed.
The standard way to quantify the degree of synchronization is via the
so called Kuramoto order parameter 
\begin{equation}
R = N^{-1}\left | \sum_{j=1}^N \mathrm{e}^{2\pi \ii \phi_j}\right | \,,
\label{eq:kura_par}
\end{equation}
where $N$ is the ensemble size and $\phi_j$, $j=1,\ldots,N$, is the proper phase 
rescaled within the unit interval.
The two above mentioned regimes correspond to: (i) $R=1$ 
(fully synchronous regime) and (ii) $R=0$ (asynchronous regime).

Besides such two extrema, partially synchronized states
may be encountered, whose universality is less clear.
Here below we introduce two major classes:
phase models (which include the Winfree model and the Kuramoto-Daido model) 
and pulse-coupled integrate-and-fire oscillators.

\subsection{Phase-oscillator models}
The dynamics of an autonomous limit-cycle oscillator is often described 
by a single equation for the phase variable. 
Without loss of generality this variable is introduced so that it evolves 
according to
\begin{equation}
\label{eq:ph1}
  \dot\phi= \nu =1/\tau\;,
\end{equation}
where $\nu$ ($\tau$) is the frequency (period) of the oscillation.
If the given oscillator weakly interacts with its environment (weakness
here means that the shape of the limit cycle is
not substantially affected by the perturbation), the phase equation modifies to
(see \cite{Kuramoto-84,Pikovsky-Rosenblum-Kurths-01} for details and further references),
\begin{equation}
\label{eq:ph3}
  \dot\phi= \nu + gQ(\phi,\psi)        \;,
\end{equation}
where $\psi$ is the phase of the forcing, $Q$ is a periodic function of both arguments, 
and $g$ quantifies the strength of the forcing or coupling.
Without loss of generality, the constant component of $Q$ can be incorporated into 
frequency $\nu$ which then becomes $g$-dependent.
In many cases $Q$ can be represented as  
\begin{equation}
\label{eq:ph4}
  Q(\phi,\psi)= \Gamma(\phi)Z(\psi)        \;,
\end{equation}
where $\Gamma(\phi)$ is the {\it phase response curve} (PRC) and
$Z(\psi)$ is the {\it forcing function}.
In globally coupled oscillators, $Z(\psi)$ can be often expressed
as the sum of the contributions of the single elements,
in which case, using the standard normalization $g\to g/N$, 
one obtains the model structure
proposed long ago by Winfree to describe biological 
rhythms~\cite{Winfree-67,Winfree-80},
\begin{equation}
\label{eq:LIF4n}
  \dot{\phi}_i= \nu + g \Gamma(\phi_i) \frac{1}{N} \sum_j S(\phi_j)\;.
\end{equation}

In the weak-coupling limit $g\ll \nu$, the interaction,
rather than being determined by the absolute phases,
is determined by phase-differences 
(see, e.g., \cite{Ermentrout-Kopell-91}). With the help of averaging
techniques, the model (\ref{eq:LIF4n}) can be indeed reduced to
the so-called Kuramoto-Daido model~\cite{Daido-92a,Daido-93a,Daido-96}
\begin{equation}
\label{eq:daido}
  \dot{\phi}_i= \nu +  \frac{g}{N} \sum_j G(\phi_i -\phi_j) \, ,
\end{equation}
identified by the single {\it coupling function}
\begin{equation}
\label{eq:conv}
  G(\xi) = \int_0^1 \Gamma(\psi+\xi)S(\psi) \dd\psi \;.
\end{equation}
A brief derivation of this known result~\cite{Golomb-Hansel-Mato-01}
is sketched in appendix \ref{app:fWtKD}.  The famous
Kuramoto-Sakaguchi model \cite{Sakaguchi-Kuramoto-86} corresponds to
$G(\xi)=\sin(-\xi+\beta)$, where $\beta=\mbox{const}$.
The structure of the Kuramoto-Daido model can be further simplified:
upon choosing a frame rotating with the common frequency $\nu$ one can get rid of the 
first term in the right hand side. Moreover, by rescaling the time variable, 
one could remove the explicit dependence on the coupling constant. 
In order to facilitate the comparison with the other models we omit such simplifications.

\subsection{The Abbott-van Vreeswijk model}

The model consists of $N$ pulse-coupled leaky integrate-and-fire (LIF) units, characterized
by the scalar variables $u_i$, $i=1,\dots N$, all restricted to the unit interval.
In the context of neural networks,
$u_i(t)$ is interpreted as the membrane potential;
it evolves according to 
\begin{equation}
\label{eq:LIF}
 \dot{u}_i(t)= a - u_i + gE(t) \;, 
\end{equation}
where $a-u_i$ represents the velocity field that is assumed to be strictly positive
(i.e. $a>1$),
while $E(t)$ is the ``mean" field arising from the interaction with the other 
oscillators and $g$ is the coupling constant. 
The evolution equation is complemented by a resetting rule:
once the potential $u_i$ reaches the threshold value $u_i=1$, it is reset to $u_i=0$, 
the neuron fires and a spike is emitted, which contributes to the generation of the field $E$.

In a globally coupled system, the field $E$ is the linear superposition 
of the pulses emitted in the past by all neurons. 
The field dynamics can be described by an additional,
linear differential equation, whose Green's function
corresponds to the pulse shape \cite{Olmi-Politi-12}. 
In the popular model of Abbott and van Vreeswijk~\cite{Abbott-vanVreeswijk-93},
the neuron firing at $t=t_0$ produces the so-called $\alpha$-pulse
whose shape is
\begin{equation}
E_\alpha(t)= \alpha^2 (t-t_0) {\rm e}^{-\alpha (t-t_0)}/N \; ,
\label{eq:alphap}
\end{equation}
where $t>t_0$, and the corresponding field equation reads
\begin{equation}
\label{eq:E}
  \ddot E(t) +2\alpha\dot E(t)+\alpha^2 E(t)= 
  \frac{\alpha^2}{N}\sum_{n|t_n<t} \delta(t-t_n) \; .
\end{equation}
From now on, the model identified by the Eqs.~(\ref{eq:LIF},\ref{eq:E}) 
will be referred to as model A.

\subsection{From the Abbott-van Vreeswijk model to phase models}
For a proper characterization of the splay state with
the help of the Kuramoto order parameter $R$, see Eq.~(\ref{eq:kura_par}), 
it is convenient to introduce phase $\phi\in [0,1)$ as
\begin{equation}
\label{eq:change2}
\phi =  -\nu \ln [1 - u/(a+g\nu)] \, ,
\end{equation}
where $\nu$ is defined by the implicit formula
\begin{equation}
\nu = -\ln^{-1}[1-1/(a+g\nu)] \, 
\label{eq:nu2}
\end{equation}
and $\phi(u=0)=0$, $\phi(u=1)=1$.
As shown in~\cite{Abbott-vanVreeswijk-93}, Eq.~(\ref{eq:LIF}) is then 
transformed to 
\begin{equation}
\label{eq:LIF4}
  \dot{\phi}_i= \nu + g \Gamma(\phi_i) \varepsilon \;,
\end{equation}
where $\varepsilon = E(t)-\nu$ and
\begin{equation}
  \Gamma(\phi) = \frac{\nu}{a+g\nu} \exp[\phi/\nu ] 
\label{eq:PRC}
\end{equation}
is the PRC. 
In this formulation the field in the asynchronous state is 
$E(t)=\nu$~\cite{Abbott-vanVreeswijk-93} and this state is 
characterized by $R=0$.
Recall that $\phi$ is taken modulo one, unless stated otherwise.

The model structure is completed by the evolution equation for the field $\varepsilon$.
Equation~(\ref{eq:E}) now becomes 
\begin{equation}
  \ddot \varepsilon +2\alpha\dot \varepsilon+\alpha^2 \varepsilon= 
  \frac{\alpha^2}{N}\sum_{n|t_n<t} \left[ \delta(t-t_n)-\nu \right] \; .
\label{eq:E2}
\end{equation}
Since the sum in the r.h.s. can be separated into contributions from $N$ neurons,
we write $\varepsilon = (1/N) \sum_j S_j$, where  
$S_j = N\sum_{i\ge 1} (E_\alpha(t-t_j^{(i)})-\nu)$ and $t_j^{(i)}$ is the 
time of the $i$th spike (counted backward starting from time $t$) emitted by
the $j$th neuron. With this representation 
we recognize a 
Winfree-type structure~(\ref{eq:LIF4n}),  with a crucial difference
in that $S_j$ cannot be expressed via the local in time value of  phase,
but has its own dynamics.

In the weak coupling limit, however, the phase of each neuron increases 
approximately linearly in time and the spikes are 
equispaced~\cite{Golomb-Hansel-Mato-01}, so that
$t-t_j^{(i)} = t-t_j^{(1)}+(i-1)/\nu = (\phi_j+i-1)/\nu$,
where $\phi_j$ is the phase of the $j$th oscillator at time $t$.
As a consequence, one can turn the explicit time dependence
of $S_j(t)$ into a phase dependence, as expected for a Winfree model.
By using the definition of $E_\alpha$ given in Eq.~(\ref{eq:alphap})
and resumming the corresponding series, one obtains
\begin{equation}
 S(\phi) = \frac{\alpha^2}{\nu} \mathrm{e}^{-\alpha \phi/\nu} 
\left[ \frac{\phi}{1-\mathrm{e}^{-\alpha/\nu}} + 
\frac{\mathrm{e}^{-\alpha/\nu}}{(1-\mathrm{e}^{-\alpha/\nu})^2} \right]
- \nu \, .
\label{eq:coup3}
\end{equation}
Eqs.~(\ref{eq:LIF4n},\ref{eq:PRC},\ref{eq:coup3}) define model B.

Next, we introduce model C: it belongs to the Kuramoto-Daido class and
is derived via averaging as an approximation of model B.
For the forcing function $S$ and the PRC given by
Eqs.~(\ref{eq:coup3},\ref{eq:PRC}), Eq.~(\ref{eq:conv}) yields the
coupling function 
\begin{equation}
\label{eq:Gphi}
G(\eta) = g_1(g_2-\eta)\mathrm{e}^{\alpha\eta/\nu}+g_3\mathrm{e}^{\eta/\nu}-g_4 \;, 
\end{equation}
see appendix~\ref{app:KDMcf} for derivation and Eq.~(\ref{eq:gcoeffs}) for 
the $g_n$ coefficients.
\begin{figure}[htb]
\includegraphics[width=0.45\textwidth,clip=true]{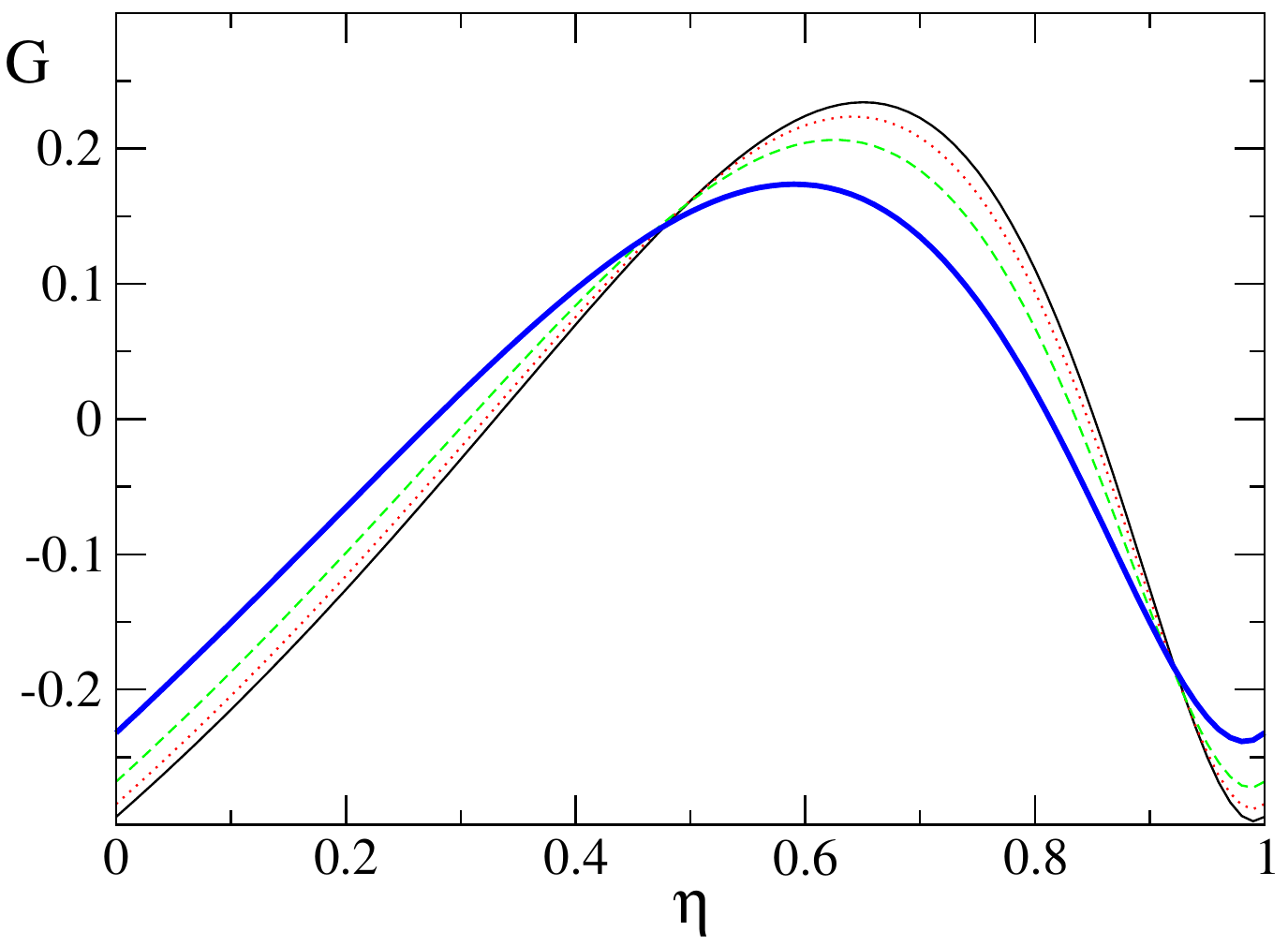}
\caption{The coupling function of model C for different values of the 
coupling strength: $g=0.02$ (solid line), $g=0.05$ (dotted line), $g=0.1$ (dashed line), 
and $g=0.2$ (bold line). 
The other parameters are $a=1.3$ and $\alpha=6$. 
The corresponding frequency values are $\nu=0.6986$, $\nu=0.7747$,
$\nu=0.7722$, and $\nu=0.8847$.
}
\label{fig1}
\end{figure}
The function $G$ is plotted in Fig.~\ref{fig1} for some
parameter values where SCPS emerges and is stable (please notice that all 
simulations below refer to $a=1.3$, while the other parameter values may vary). 
The coupling function $G$ does not reveal any special structure: 
it has one maximum and one minimum within the period. It can be checked, that 
$G(0)=G(1)$; however, $G'(0)\ne G'(1)$.
The implications of such properties are extensively discussed in the next section.

\subsection{Back to pulse coupled oscillators: a computationally efficient model}
As a corollary of the previous analysis, Winfree-type 
models characterized by different phase-response curves $\Gamma$ and 
different forcing function $S$, but identical convolution products $G$ (see Eq.~(\ref{eq:conv})) 
are expected to be equivalent. 
Among them, it is instructive to consider the model with a $\delta$-like forcing function and 
$\Gamma(\phi)=G(\phi)$,
\begin{equation}
\label{eq:newpul}
  \dot{\phi}_i= \nu + \frac{g}{N} G(\phi_i) \sum_j (\delta (\phi_j)-1) ,
\end{equation}
where we have subtracted $1$ to ensure a zero average of the forcing function
like in the original setup. 
As expected for a Winfree-type model, the argument of 
the $\delta$-function here is the phase. It can be transformed into a
time-dependent function by substituting $\phi/\nu \to t$ into the argument of the
$\delta$-function
\begin{equation}
\label{eq:newpul2}
  \dot{\phi}_i= \nu - gG(\phi_i) + \frac{g}{N\nu} G(\phi_i) \sum_j \delta (t-t_1) ,
\end{equation}
where $t_1$ is the time when any oscillator is reaching the threshold $\phi =1 $
This is a standard model of $\delta$-coupled oscillators with a
weakly phase-dependent velocity field. In the following we shall refer to it
as to model D.

From a computational point of view it is preferable to change variables,
introducing $\theta_i$, according to 
\begin{equation}
\label{eq:newang}
  \frac{\dd\phi_i}{\dd\theta_i}= R(\phi_i) \equiv \frac{\nu - g G(\phi_i)}{\nu_0}\;,
\end{equation}
so that $\dot \theta_i = \nu_0$ (with a further
adjustment of the PRC that has to be divided by $R(\phi_i)$), while the interaction
terms would still be easy-to-handle $\delta$-spikes. 
In fact, since the time derivative between the spikes is constant, the simulation of this model 
does not require a differential-equation solver and can be performed very efficiently.
The price to pay is that $\theta$ is no longer appropriate to characterize the splay state,
as the corresponding Kuramoto-order parameter would now differ from zero.

Finally, it is necessary
to comment about a subtle point: since the PRC is negative for $\phi = 0$,
the effect of an incoming spike on the $i$th neuron whose phase is just above zero
may push it backward below zero. If one interprets $\phi$ as a true phase,
this would mean that the $i$th neuron is set below threshold and thus ready 
to fire again, a phenomenon that does not happen in the original formulation of the model. 
We should in fact interpret $\phi$ as the membrane potential $u$ in
Eq.~(\ref{eq:LIF}) and avoid the identification of $\phi<0$ with $1+\phi$.

\section{Model comparison}
\subsection{Splay state}

The splay state and, more precisely, its stability is the first ground
where the three models can be compared.
The stability analysis is performed in the thermodynamic limit
$N\to\infty$ by introducing 
the probability distribution $P(\phi,t)$ of the phases and 
writing the continuity equation 
\begin{equation}
\frac{\partial P}{\partial t} = -\frac{\partial (\dot \phi P)}{\partial \phi} = -\frac{\partial J}{\partial \phi}\;,
\label{eq:cont}
\end{equation}
where $J$ is the corresponding current. 

The three models require different approaches:
for instance in model A it is necessary to include the field dynamics into the analysis, 
while model C does not require any perturbative expansion. 
In all three cases, however, in the small-$g$ limit the relevant eigenvalues
can be expressed as (see appendix~\ref{sec:linsplay} for a detailed account of the calculations),
\begin{equation}
\mu_n = 2 \pi \ii n \nu +g\delta_n \;,
\label{eq:eigenvdel}
\end{equation}
where
\begin{equation}
\delta_n =  \left[ \frac{\alpha^2\nu^2 }{a+g\nu} \right]
 \frac{\mathrm{e}^{1/\nu} -1}
{ (\alpha+2\pi \ii n\nu)^2 (1+2\pi \ii n\nu)}  \;,
\label{eq:eigenvdel2}
\end{equation}
while the corresponding eigenvectors are Fourier modes of increasing frequency.
This result reveals a perfect correspondence among the three models in the weak-coupling limit. 

In particular, it is interesting to notice that the splay state becomes unstable (along the
direction identified by the first Fourier mode) if $\alpha$ exceeds the critical  value
\begin{equation}
\alpha_c = -1 + \sqrt{1+4\pi^2\nu^2} \, .
\label{crit0}
\end{equation}
The loss of stability in model A for $g=0.3$ was discovered in Ref.~\cite{vanVreeswijk-96},
where it was shown that it corresponds to the onset of SCPS (see below).
Our analysis reveals that this critical phenomenon extends down to the weak coupling limit
and is therefore more general than initially believed.

\begin{figure}
\includegraphics[width=0.45\textwidth,clip=true]{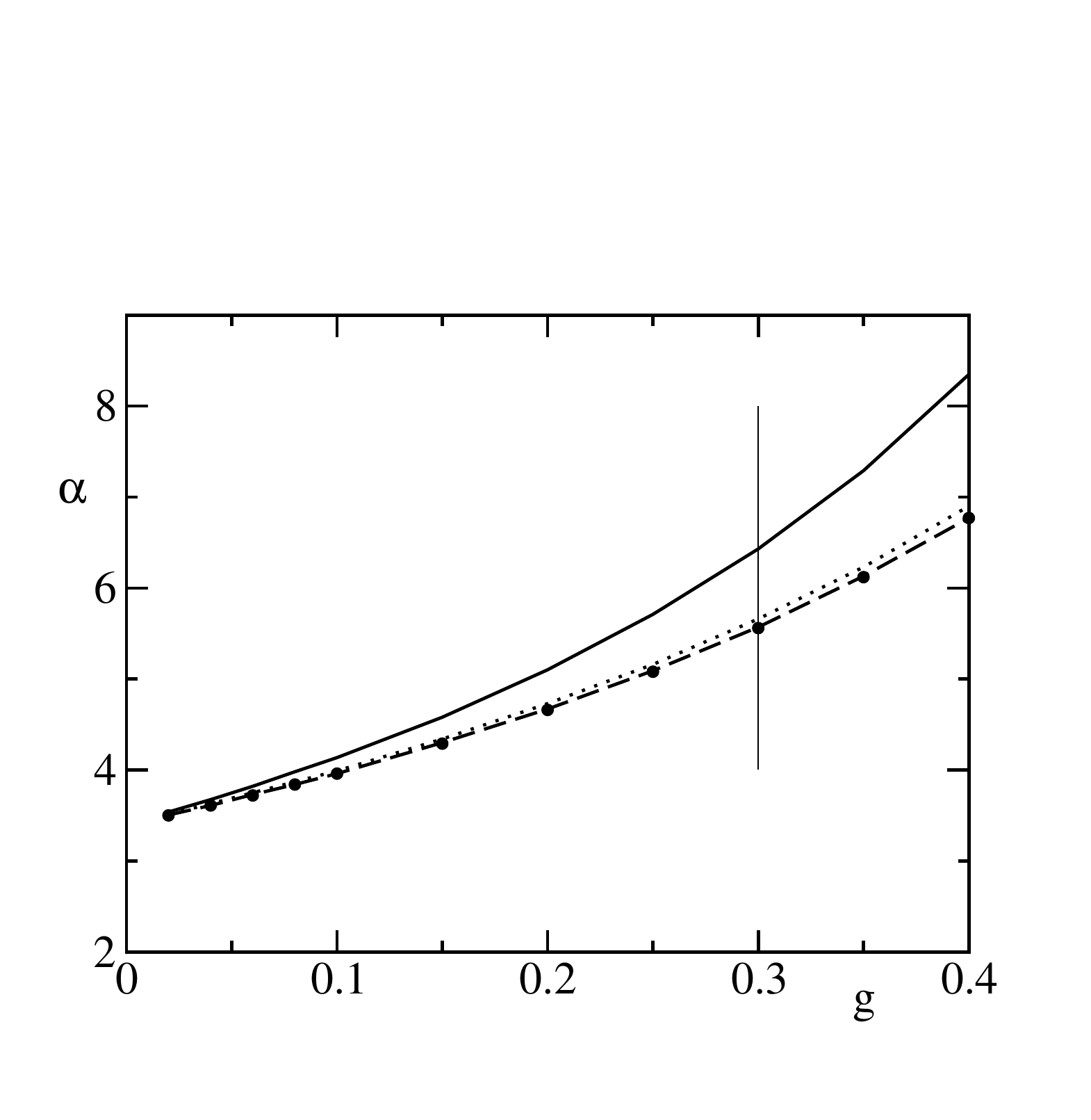}
\caption{
Loss of splay state stability in models A (solid curve), 
B (dashed curve),  C (filled circles), and D (dotted curve). 
The curves for models A,B,D are 
obtained by numerical study of the correspondent model for $N=200$ oscillators.
The curve C corresponds to the perturbative calculations, see Eq.~(\ref{crit0}).
The vertical line identifies the locus of points numerically analyzed 
in Ref.~\cite{vanVreeswijk-96}.}
\label{fig:lifdiag}
\end{figure}

In Fig.~\ref{fig:lifdiag} we report the bifurcation diagram in the plane $(g,\alpha)$, 
for $a=1.3$. 
The solid curve, obtained by simulating model A for large systems, 
separates the lower region, where the splay state is stable from the upper one, where SCPS is observed. 
The vertical straight line at $g=0.3$ corresponds to the interval of $\alpha$-values
investigated by van Vreeswijk. 
The dashed curve corresponds to the perturbative result (\ref{crit0}) as well as to the
transition line of model C: it provides an 
excellent approximation even for relatively large $g$ values.

Quite surprisingly, numerical estimates of the transition line for
model B do not reveal appreciable deviations from the perturbative
prediction, suggesting that higher order
terms are almost negligible in the Winfree setup (at least up to $g\approx 1$).
The same agreement is observed for the $\delta$-coupled oscillators in model $D$.

\subsection{Synchronous state}
While considering the synchronous regime, it is instructive to monitor not only stability but 
also the period $T$ of the solution as, contrary to the previous case, it is affected by the 
coupling strength. 
Let $\tau$ be the period of the uncoupled system.
As follows from Eq.~(\ref{eq:nu2}) for $g=0$, $\tau=-\ln(1-1/a)$. 

For model A, by making use of some general formulas derived in \cite{Olmi-Politi-Torcini-10} 
it is found (see appendix \ref{sec:linsyn}) that in the weak-coupling limit the period 
can be written as $T=\tau+\delta T$, where
\begin{equation}
\delta T =  \frac{g \tau\alpha^2}{a} H\;,
\label{eq:LIFtau}
\end{equation}
where
\begin{equation}
H =  \frac{ \mathrm{e}^{-\alpha\tau}(\mathrm{e}^\tau-1)}
{(\alpha-1) (1-\mathrm{e}^{-\alpha \tau})^2} 
-\frac{\nu(1\!-\!\mathrm{e}^{-(\alpha-1)\tau})}{(\alpha-1)^2
(1-\mathrm{e}^{-\alpha \tau})} \;.
\label{eq:B}
\end{equation}
For models B and C it is instead found that (see again appendix \ref{sec:linsyn}) 
\begin{equation}
\label{eq:wintau}
  \delta T = \frac{g\tau \alpha^2}{a}  \left [H + \frac{\nu^2}{\alpha^2}
(\mathrm{e}^{\tau} -1) 
\right ] \;.
\end{equation}
These expressions indicate that the agreement between the original LIF setup 
and Winfree and Kuramoto-Daido-type models is not perfect: a difference manifests itself
already at the first order in $g$, i.e. 
\[
\frac{\delta T_{B,C}-\delta T_A}{\tau}=\frac{g}{a\tau^2}(e^\tau-1)\approx 1.2 g\;.
\]
Although the discrepancy is not small, it is more on a quantitative 
than on a qualitative level. 

\begin{figure}[htb]
\includegraphics[width=0.45\textwidth,clip=true]{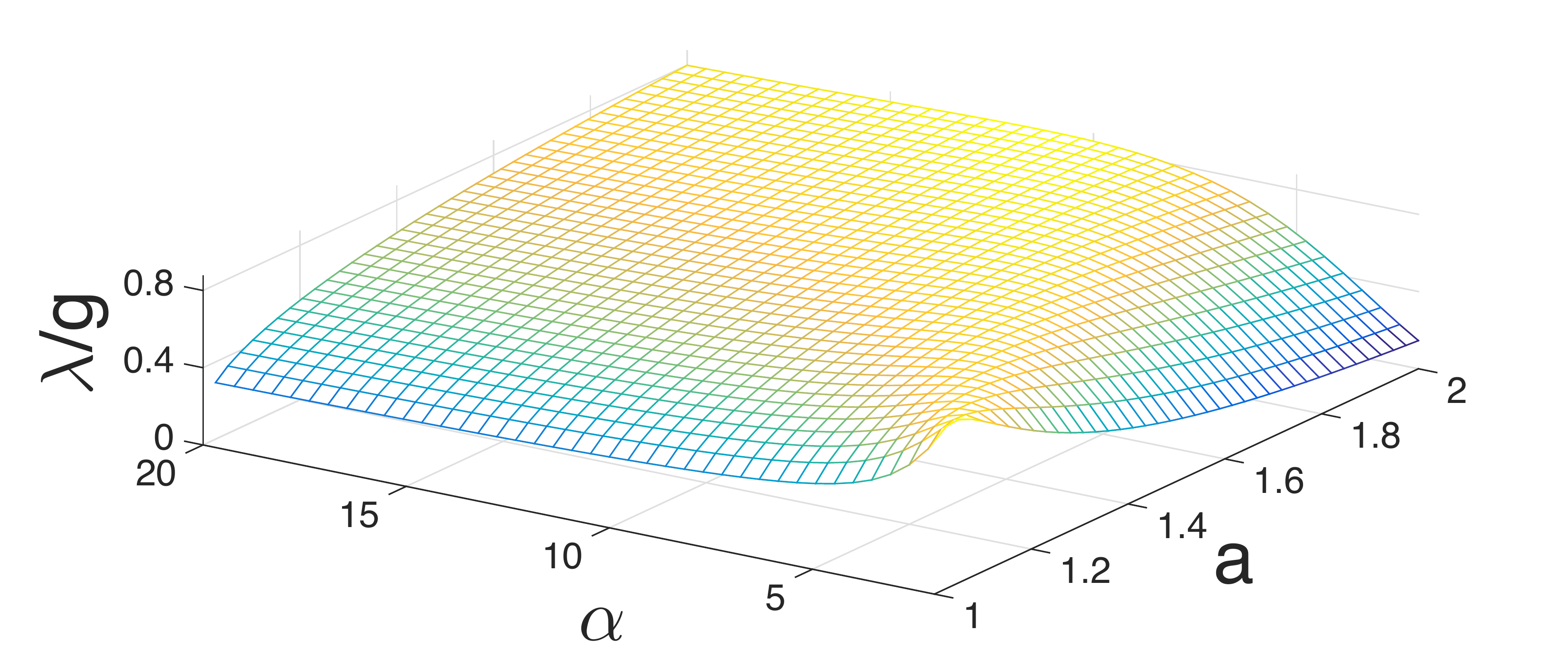}
\caption{The ratio of the Lyapunov exponent and coupling strength, 
$\lambda/g$, versus $\alpha$ and $a$ for
the synchronous solution of models A, B, see Eq.~(\ref{eq:LIFsy3}).
}
\label{fig:lyapsyn}
\end{figure}

The stability analysis of the synchronous solution for models A and B (for $g\ll 1$), 
(again performed in appendix \ref{sec:linsyn}) yields the Lyapunov exponent
\begin{equation}
\label{eq:LIFsy3}
\lambda = -g \frac{\alpha^2}{a} \left[ 
\frac{ \alpha \mathrm{e}^{-\alpha\tau}(\mathrm{e}^\tau-1)}
{(\alpha-1) (1-\mathrm{e}^{-\alpha \tau})^2} 
-\frac{\nu(1\!-\!\mathrm{e}^{-(\alpha-1)\tau})}{(\alpha-1)^2
(1-\mathrm{e}^{-\alpha \tau})} \right]\;.
\end{equation}
For $\alpha>1$ and $g>0$ the synchronous solution turns out to be unstable,
as it can be appreciated in Fig.~\ref{fig:lyapsyn}.

As for the model C, the stability of its synchronous solution is 
given by $\lambda = g G'(0)$, where $G'(0)$ is the derivative
in the origin (see the appendix \ref{sec:linsyn}): here it arises an additional difference.
The point $\phi=0$ is to be identified with $\phi=1$, but the derivative of $G(\phi)$ in the two
points is different: in practice, this means that the right derivative differs from the
left one; Eq.~(\ref{eq:LIFsy3}) corresponds to the right derivative.
The difference between the two derivatives is however somehow irrelevant,
as it does not affect the sign (at least for our selection of the PRC and pulse shape).

Thus, the perturbative analysis 
shows that in the limit $g\ll 1$, $\lambda$ the Winfree and Kuramoto-Daido models are almost 
but not perfectly equivalent to the LIF model: the leading correction for the period of the synchronous 
regime differ in models B and C.

\subsection{Partial synchronization}

Self-consistent partial synchronization has been observed only in a few setups,
but the stability analysis of the splay state discussed above in this section
suggests that this phenomenon might be more general than so far believed.
In fact, here we show that SCPS arises in all A-D models and it can be
analyzed perturbatively in the weak-coupling limit.

\begin{figure}
\includegraphics[width=0.45\textwidth,clip=true]{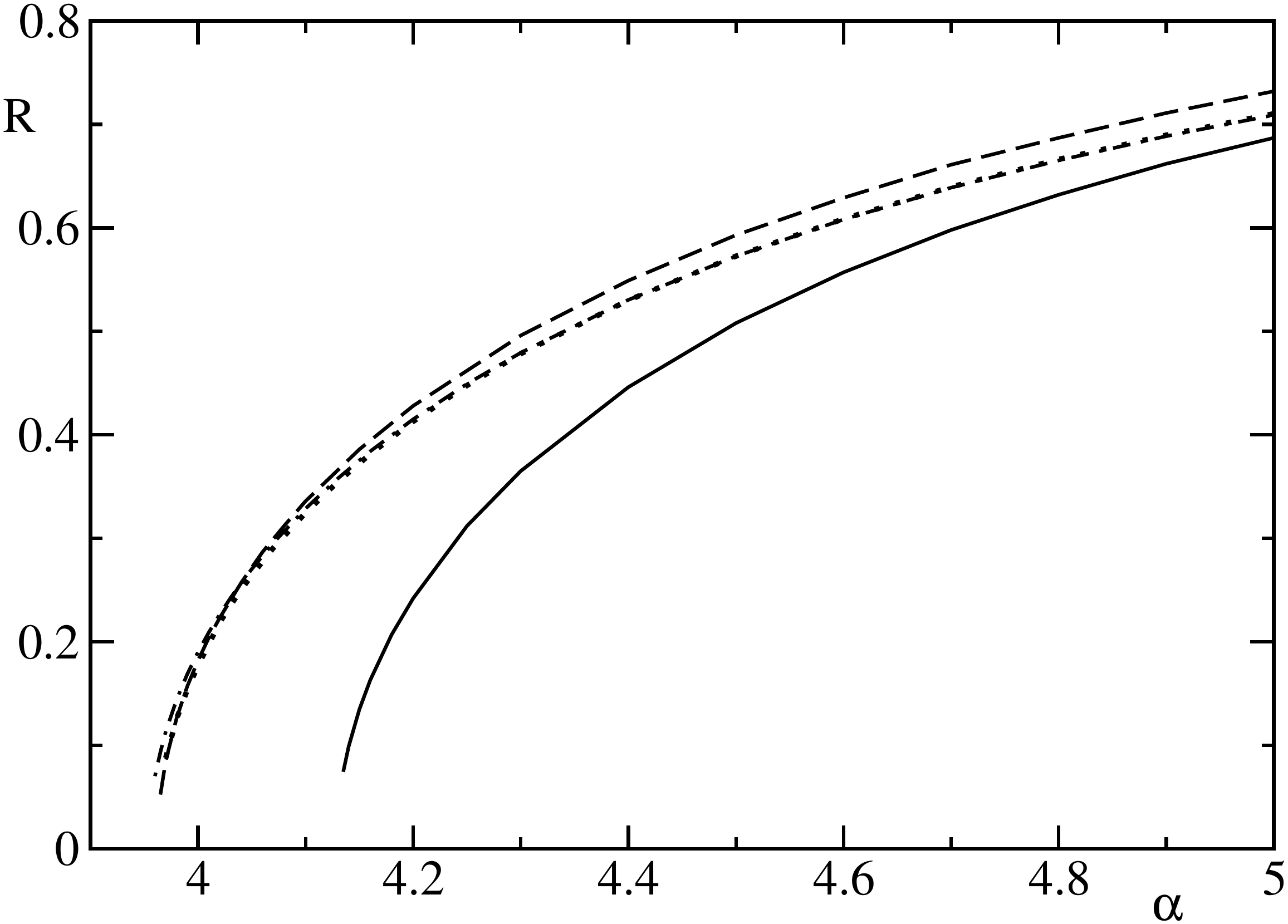}
\caption{
Average value of the Kuramoto order parameter for $g=0.1$, and $a=1.3$:
the solid, dashed, dotted, and dotted-dashed lines refer to models A, B, C, and
D, respectively.}
\label{fig:graph_ord}
\end{figure}

A way to spot SCPS is via a nonzero value of the
Kuramoto order parameter $R$. In Fig.~\ref{fig:graph_ord} it can be seen
that a transition towards such a regime occurs when the inverse pulse-width $\alpha$ is
increased.
The curves obtained for the four models are rather close to each other, confirming an
agreement that is expected from the perturbative analysis of the splay state.
The more sizable deviations concern model A, suggesting that
the field dynamics is not entirely negligible.
Quite remarkably, the outcome of model D is also consistent
(see the dotted-dashed curve in Fig.~\ref{fig:graph_ord}),
confirming that the effect of a smooth pulse shape can be harmlessly
transferred to the PRC.

Let us now identify a signature of SCPS:
a difference between the average frequency $\w$ of the oscillators 
(the same for all of them) and the frequency of the mean field
\begin{equation}
\W=\langle \dot\Theta \rangle\;, \quad\mbox{where}\quad
\Theta=\mbox{arg}\left (N^{-1}\sum_j\mathrm{e}^{\ii \phi_j}\right )\; ,
\end{equation}
where $\langle\cdot\rangle$ means time average.

\begin{figure}
\includegraphics[width=0.45\textwidth,clip=true]{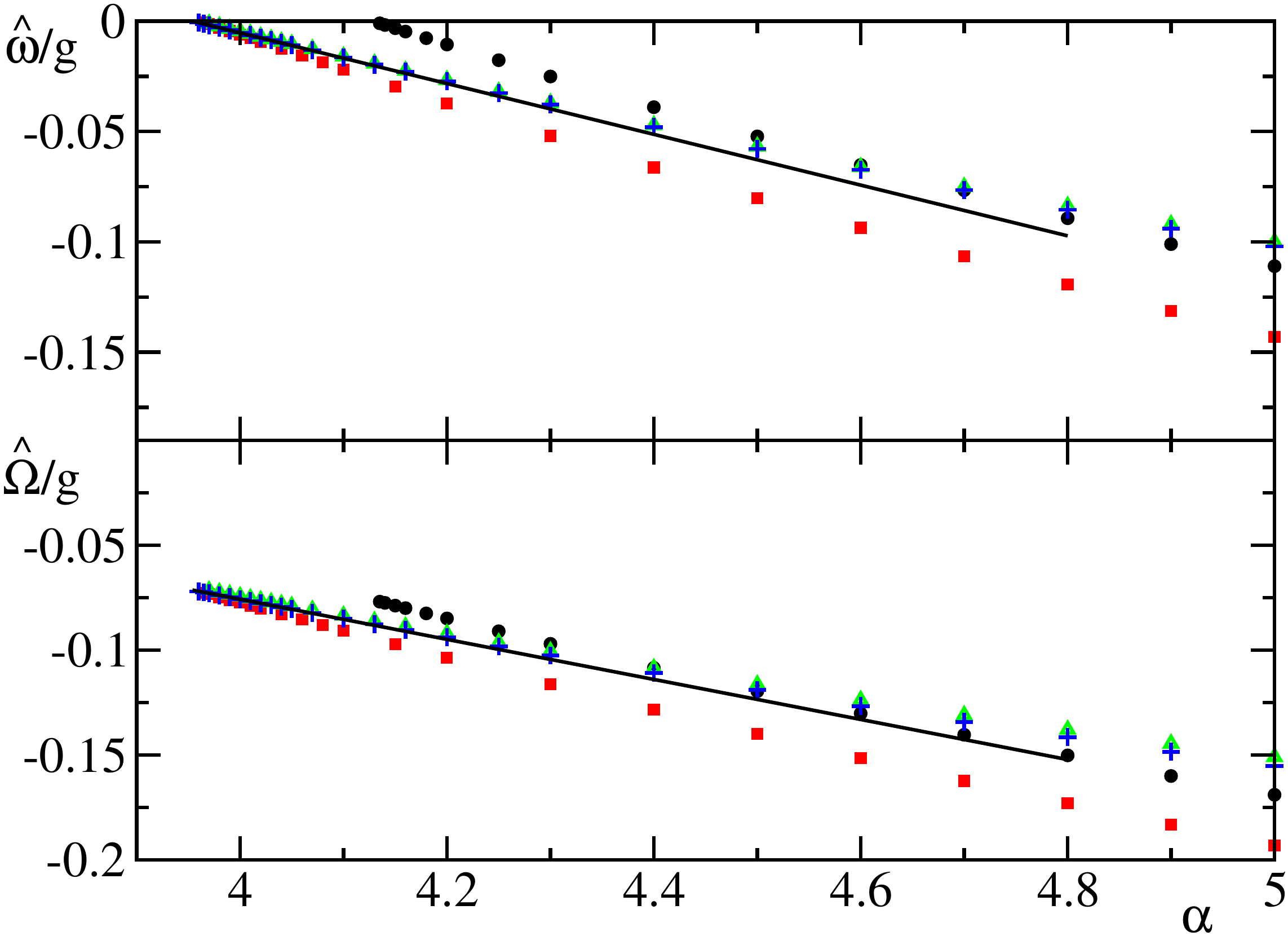}
\caption{(Color online) Average frequencies of the oscillators (top panel) and of the mean field 
vs $\alpha$; for $g=0.1$ and $a=1.3$.
In both cases, the frequency of the splay
state is subtracted, see Eq.~(\ref{eq:relfreq}), and the result is scaled with respect 
to the coupling strength $g$. Black circles, red squares, blue pluses, 
and green triangles correspond
to models A, B, C, and D respectively. 
The two solid curves are the outcome of the perturbative calculations carried 
out with model C (see Sec.~\ref{sec:KuraDaido}).}
\label{fig:graph_freq}
\end{figure}

The results are plotted in Fig.~\ref{fig:graph_freq} (for the
same parameter values as in Fig.~\ref{fig:graph_ord}).
The two frequencies are reported after
subtracting the bare frequency $\nu$ of the splay state to better
appreciate the implication of the transition; i.e. we plot the relative frequencies
\begin{equation}
\hat\w=\w-\nu\;,\quad \hat\W=\W-\nu\;.
\label{eq:relfreq}
\end{equation}

In the upper panel we see that the oscillator frequency $\hat\omega$ vanishes at the critical
$\alpha$-value below which SCPS disappears. 
All curves lie below zero: this means that in 
SCPS the oscillators are slower than in the splay state.
In the lower panel, one can see that the mean field frequency $\hat\W$ 
is smaller than that of the oscillators:
this is a typical signature of SCPS:
it means that the oscillators ``move" faster than their distribution.
(Cf. with the results for the nonlinear Kuramoto-Sakaguchi-like model
in Refs.~\cite{Rosenblum-Pikovsky-07,*Pikovsky-Rosenblum-09}, where 
the oscillators can have any frequency relative to the mean field.) 
At the transition, the value of $\hat\W$ coincides with 
the frequency of the Hopf bifurcation. Once again, one can notice a
similar kind of agreement among the three models.

\begin{figure}
\includegraphics[width=0.45\textwidth,clip=true]{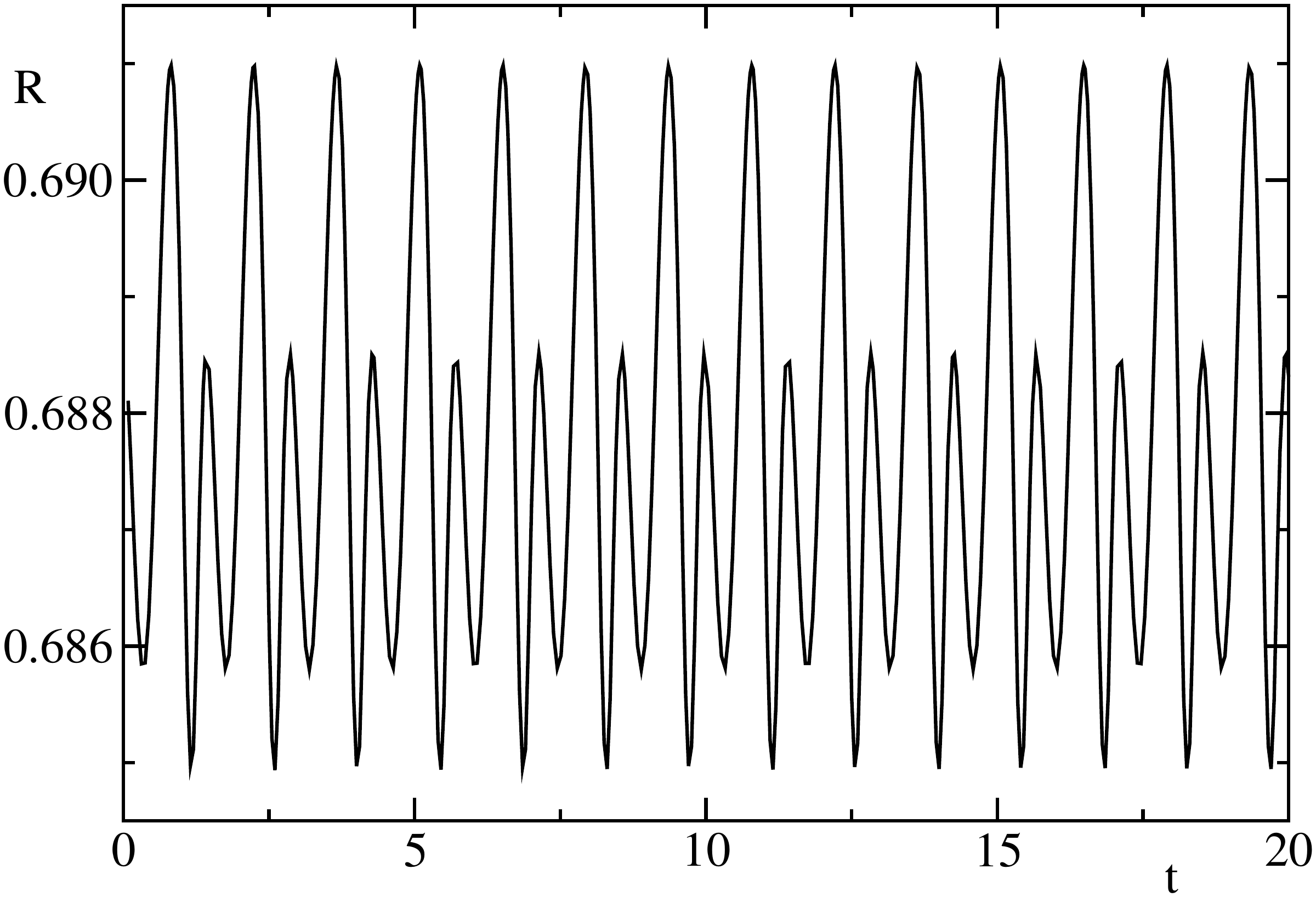}
\caption{
Evolution of the Kuramoto order parameter in model A
for $g=0.1$, $\alpha=5$ and $a=1.3$.}
\label{fig:liford}
\end{figure}

Finally, we plot in Fig.~\ref{fig:liford} the time trace of the
Kuramoto order parameter $R$ for the model A and an $\alpha$-value above
threshold. There, one can see small periodic oscillations, which are still
present in model B (data not shown), but completely absent in model C. 
As explained in the next section, this behavior is a consequence of
the invariance of the evolution equations under a phase shift.

\section{Partial synchronizaton: a perturbative approach}
\label{sec:KuraDaido}

Within the Kuramoto-Daido setup, the forces depend on phase differences.
Accordingly, there may exist non-uniform phase distributions that move
rigidly in time. They can be viewed as fixed points of Eq.~(\ref{eq:cont}) 
in a suitably moving frame. The first example of such a regime was perhaps 
discussed in \cite{Kuramoto-91}, where the author developed an approximate 
description of the LIF model in the presence of delayed pulses. 
Here below we show that such states, sometimes referred to as rotating 
waves \cite{Hansel-Mato-Meunier-95}, are instances of SCPS.
The representation of SCPS as a fixed point allows developing a perturbative
approach and thereby deriving approximate analytical expressions to be
compared with the numerics.

Let us start expressing Eq.~(\ref{eq:cont}) in a frame that rotates with the
(yet unkwnown) frequency $\W$, by mapping $\phi\to\phi-\W t$, and then
set $\frac{\partial P}{\partial t}=0$. By assuming that the velocity field
is defined as in Eq.~(\ref{eq:daido}) (for $N\to\infty$), one obtains
\begin{equation}
\frac{\partial}{\partial \phi}
\left[ \left(-\hat\W + g\int d\psi G(\phi-\psi)P(\psi)\right) P(\phi) \right]=0\;,
\label{eq:partialKD}
\end{equation}
where $\hat \W$ is an unknown quantity, to be determined self-consistently.
Upon integrating the above equation,
\begin{equation}
\left[ -\hat\W + g\int d\psi G(\phi-\psi)P(\psi)\right] P(\phi) = \eta=\mbox{const} \;,
\label{eq:partialfix}
\end{equation}
where the probability flux $\eta$ is also to be determined.
Since phases are rescaled to the unit interval, the flux $\eta$
corresponds to the difference between the average frequency of the oscillators 
and that of the mean field,
\begin{equation}
\eta=\hat\w-\hat\W=\w-\W\;.
\end{equation}
In general, there maybe two classes of solutions of Eq.~(\ref{eq:partialfix}), 
characterized by $\eta=0$ and $\eta\ne 0$, respectively. 
In the former case, the expression in square brackets
must vanish. 
By going in Fourier space, it can be easily seen that no such probability 
distribution can satisfy the condition if all Fourier components $\tilde G_n\ne 0$. 
On the other hand, whenever
$\tilde G_n=0$, $\tilde P_n$ is allowed to be different from zero. 
Such distributions are just
marginally stable and any arbitrarily small amount of noise would smooth them out.
The only physically interesting solutions are those of the second class.

\begin{figure}
\includegraphics[width=0.45\textwidth,clip=true]{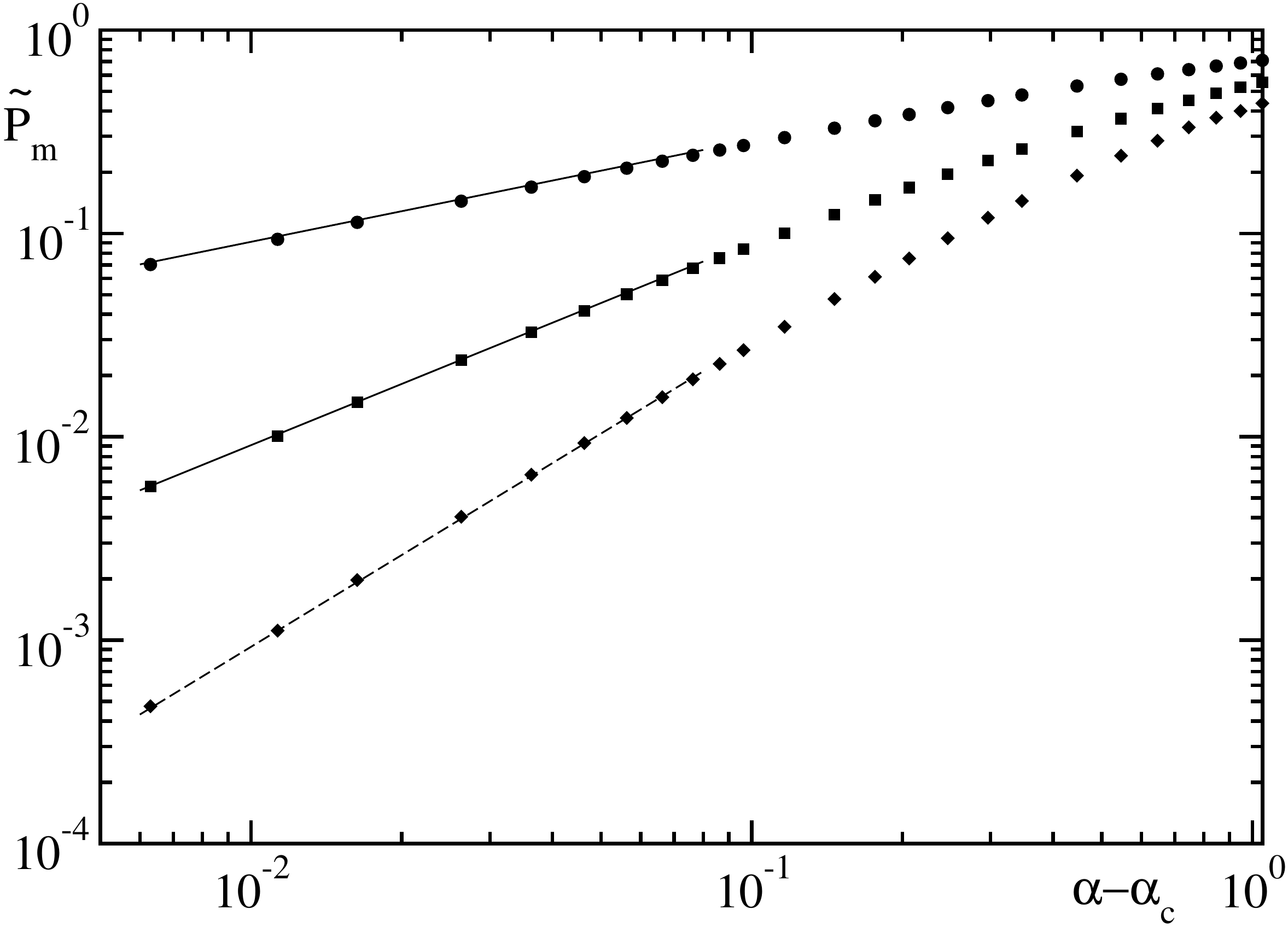}
\caption{First Fourier components of the phase-distribution for $g=0.1$ in model C
versus the distance from the critical point.
Circles, squares and diamonds refer to $m=1$, 2, and 3, respectively.
$|\tilde P_1|$ corresponds to the Kuramoto order parameter, see Eq.~(\ref{eq:kura_par}). 
The two solid curves
(which scale with exponents 1/2, 1, respectively) are the outcome of the analytic
calculation. The dashed curve is the outcome of a best fit with a slope 3/2.}
\label{fig:scalrho}
\end{figure}

Determining $P(\phi)$ is not an easy task.  Let us start discussing the parameter
region close to the bifurcation point, where deviations from a flat distribution
are small.
It is convenient to rewrite Eq.~(\ref{eq:partialfix}) in Fourier space, 
\begin{equation}
\hat\W \sum_n \tilde P_n \mathrm{e}^{-2\pi \ii n \phi} -g
 \sum_{m,n} \tilde G_m \tilde P_m \tilde P_n \mathrm{e}^{-2\pi \ii (m+n) \phi} =\eta \;,
\label{eq:fixfou}
\end{equation}
and to decompose it into equations for the single components, obtaining
\begin{equation}
\hat\W \tilde P_k - g\sum_{m} \tilde G_m \tilde P_m \tilde P_{k-m} =\eta\delta_{k0} \;.
\label{eq:fixfou2}
\end{equation}
The simulations reported in Fig.~\ref{fig:scalrho} suggest that higher order
harmonics are increasingly negligible upon approaching the bifurcation.
Therefore, we restrict the 
analysis to the modes $k=1$ and $k=2$
(notice that $\tilde P_0=1$ for normalization reasons, while $\tilde G_0=0$
by definition, since the constant term of the coupling function is absorbed into
the frequency).
From the equation for the mode $k=0$ we obtain
\begin{equation}
\tilde G_1^r |\tilde P_1|^2 + \tilde G_2^r |\tilde P_2|^2 = (\hat\W-\eta)/2g \, ,
\label{eq:fixfou5}
\end{equation}
where the superscript $r$ means that the real part is being considered. 
For $k=1$ and $k=2$ we find,
\begin{align}
&\tilde G_1 + \tilde G_2 \tilde P_2  + \tilde G_1^* \tilde P_{2} =\hat \Omega/g \;,
\label{eq:fixfou6}\\
&\tilde G_1 \tilde P_1^2 + (\tilde G_2-\hat\Omega/g) \tilde P_2 = 0 \;,
\label{eq:fixfou7}
\end{align}
where we have assumed (without loss of generality) that $\tilde P_1$ is real
(the phase of the solution is arbitrary and we can set the origin
as we prefer).

Let us now imagine that upon variation of the control parameter $\mu$,
there exists a transition to SCPS for $\mu=\mu_c$.
Since $\tilde P_2=0$ at the transition, from Eq.~(\ref{eq:fixfou6}) it 
follows that, $g\tilde G_1(\mu_c)=\hat\W$; 
we call this specific value $\hat\W_0$. Therefore, slightly above the threshold,
$g\tilde G_1 = \hat\W_0 + g\tilde G'_1 \delta \mu$ and
$\hat\W = \hat\W_0 + \delta \W$, where the prime denotes the derivative with respect
to $\mu$, while $\delta \W$ has to be determined.
A solution of Eq.~(\ref{eq:fixfou6}) is, to the leading order,
\begin{equation}
\tilde P_2 = \frac{\delta \W - g\tilde G'_1\delta \mu}{g\tilde G_2 +\hat\W_0} \;,
\label{eq:fixfou8}
\end{equation}
so that now Eq.~(\ref{eq:fixfou7}) yields $\tilde P_1^2$. Next, using that 
$\tilde P_1$ (and thus $\tilde P_1^2$) is real, 
we obtain $\delta \W$ from the condition $\mbox{Im}(\tilde P_1^2)=0$, 
see appendix~\ref{sec:app4} for details.
As a result, we find that $\delta\W\sim\delta\mu$, $\tilde P_1^2\sim\delta \mu$.
A physically meaningful solution $\tilde P_1\sim\sqrt{\delta \mu}$ 
exists for $\delta\mu>0$, i.e. above the bifurcation point, 
and Eq.~(\ref{eq:fixfou7})
implies that $\tilde P_2$ grows linearly and is in general complex,
meaning that it is shifted with respect to the phase of $\tilde P_1$.
Finally, neglecting the term proportional to $P_2^2$ in Eq.~(\ref{eq:fixfou5}), 
we determine the last unknown, $\eta$,
\begin{equation}
 \hat\W-\eta=  2g\tilde G_1^r |\tilde P_1|^2 \;.
\label{eq:Kexp}
\end{equation}
Notice that both $\tilde\W$ and the frequency difference $ \hat\W-\eta$  
depend linearly on the control parameter 
in the vicinity of the bifurcation.

These perturbative results  can be compared with the numerical simulations illustrated
in the previous section: $\alpha$ plays the role of the control parameter $\mu$.
By computing $\tilde P_1$ and $\tilde P_2$ for $a=1.3$ and $g=0.1$
(see  Appendix~\ref{sec:app4}), one obtains the data reported 
in Fig.~\ref{fig:graph_freq}. The two frequencies $\hat\omega$ and $\hat\W$ 
reveal an excellent agreement with the direct simulation of the three models.
Moreover, in Fig.~\ref{fig:scalrho}, one can  see that the
theoretical results (see the two upper solid lines) reproduce 
perfectly the behavior of the first two Fourier modes of the
phase distribution. 

\begin{figure}
\includegraphics[width=0.45\textwidth,clip=true]{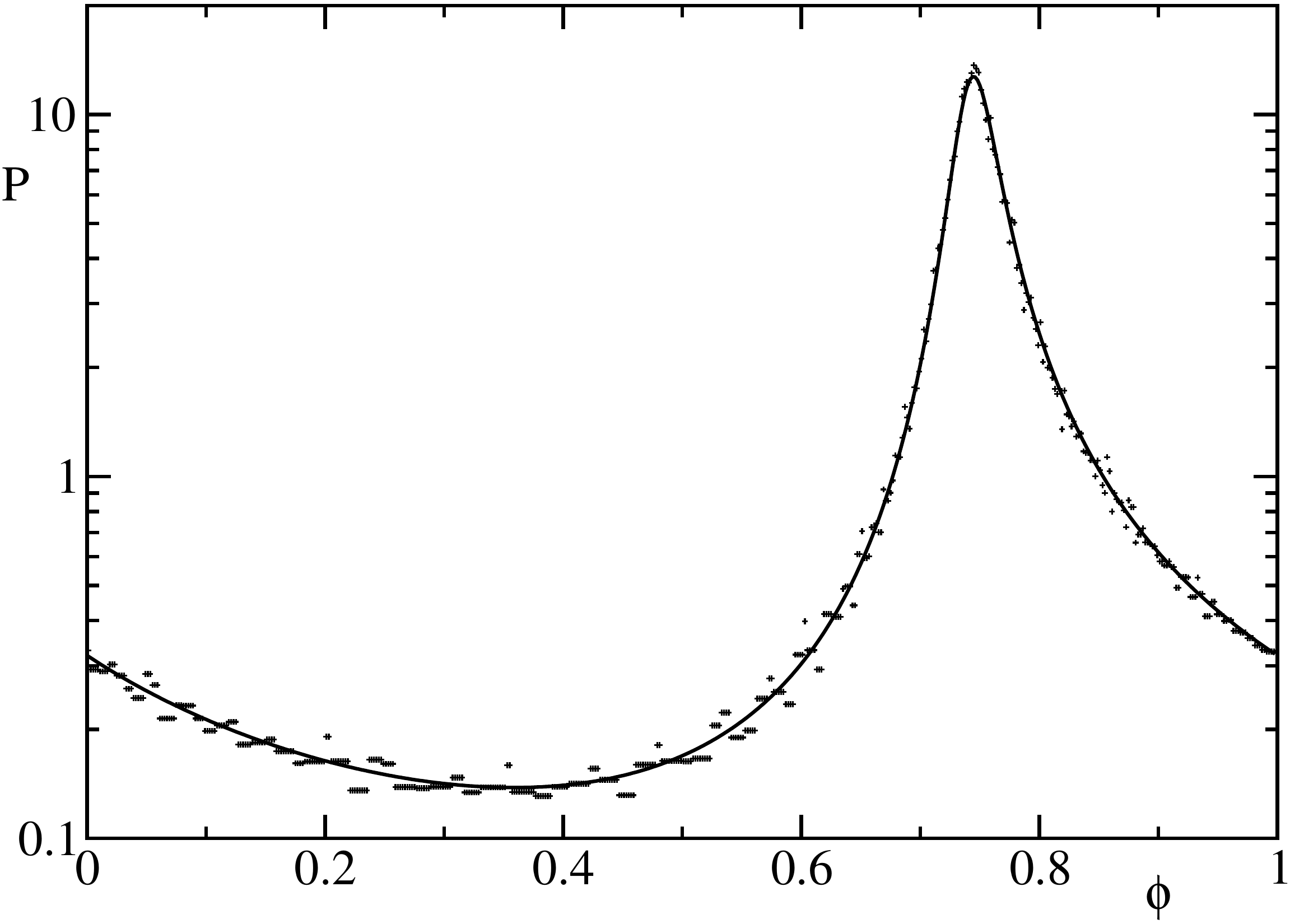}
\caption{Probability distribution in the partial synchronization regime for $\alpha=4.7$
for the model C. The solid curve is the outcome of the recursive procedure discussed
in the text, while the pluses refer to direct numerical simulations with $N=1000$.
}
\label{fig:compa}
\end{figure}

Away from criticality, many Fourier modes come into play and a perturbative scheme
is no longer effective. The distribution $P(\phi)$ can be nevertheless accurately
determined by interpreting Eq.~(\ref{eq:partialfix}) as the fixed point of the
recursive relation
\begin{equation}
P_{n+1}(\phi) = \frac{\eta}{g\int  G(\phi-\psi)P_n(\psi)\dd\psi -\hat\W} \, .
\label{eq:partialfix2}
\end{equation}
This equation shows that $\eta$ can be determined by imposing the normalization of the r.h.s.. 
Numerical studies have revealed that generically the recursive
procedure either converges to the flat distribution $P(\phi)=1$ or develops
nonphysical negative values. We have found that upon
tuning $\hat\W$, one can pass from the former to the latter regime, that are
separated by a critical $\hat\W$ value for which the recursive procedure converges
to a given shape with some shift. 
Upon changing the initial distribution, different phase shifts may be found:
the correct solution is the one characterized by a zero shift (a true fixed point). 
Luckily, this objective can be reached by controlling a single parameter 
of the initial distribution:
we have found that the most effective one, is the width of the distribution itself.
Altogether, in spite of the fact that the fixed point is a infinite-dimensional
function, its shape can be determined by tuning two parameters only.
The outcome of this procedure is shown in Fig.~\ref{fig:compa} for $\alpha=4.7$. 

\section{Summary and open problems}

In this paper we have performed a quantitative comparison 
of different model-classes of (phase) oscillators.
A perturbative analysis of integrate-and-fire oscillators and of the corresponding
Winfree and Kuramoto-Daido models reveals a substantial equivalence.
The stability of the splay state is perfectly reproduced: the whole
spectrum of eigenvalues coincides for all of the three models up to leading order.
As for the synchronous solution, the leading correction to its frequency 
in the Winfree and the Kuramoto-Daido models differs from that found in the LIF model.
Moreover, the Kuramoto-Daido model fails to reproduce its stability (left
stability differs from the right stability as a consequence of a
nonanaliticity in the coupling function), although
the difference is quantitative, but not qualitative.

The comparison has been extended to the SCPS regime which arises from
the splay state through a Hopf bifurcation. In this case, a mostly numerical
analysis reveals again an excellent agreement among the various models.
The largest deviations are observed for the LIF model, signaling that
the field dynamics is not entirely negligible even in the small coupling limit.

An important consequence of our comparative studies is the overall evidence
that SCPS is not specific of integrate-and-fire oscillators, but rather
universal, instead.
In particular, it is not necessary to invoke a dependence on 
the order parameter, as assumed in~\cite{Rosenblum-Pikovsky-07,*Pikovsky-Rosenblum-09}.

Furthermore, the mapping of the original LIF dynamics onto a Kuramoto-Daido-type model 
has offered the opportunity to develop a perturbative treatment of SCPS. 
In fact, in such a setup, SCPS
corresponds to a uniform rotation of the probability density that can be
seen as a fixed point in a suitably moving frame and thereby analysed with
powerful techniques.

The actual observation of SCPS in a Kuramoto-Daido setup such as model C 
opens the question of identifying the minimal requisites for its observability.
If the coupling function is composed of only one harmonics (the Kuramoto-Sakaguchi model),
it is known that something similar to SCPS can be observed only in the special case of 
phase shift equal to $\pi/2$, where it is anyhow marginally stable.
In a separate publication we will show that it is sufficient to add a second harmonic
to observe a stable and robust regime of self-consistent partial synchronization.

Finally, the good correspondence between model D and the other
phase models implies that restricting the study to $\delta$-coupled 
integrate-and-fire oscillators is not a true limitation in so far as 
finite pulse widths can be reduced to such a class by suitably adjusting
the phase response curve.
Such an equivalence has practical advantages, as the former class of models
is easier to simulate.

To what extent the correspondence among the models extends to large coupling strengths is
also not known: this is another point that is worth to investigate in the future.

\section*{Acknowledgment}
AP wishes to acknowledge the von Humboldt Foundation for the financial
support, which made the collaboration possible.

\appendix

\section{From Winfree to Kuramoto-Daido}
\label{app:fWtKD}

In the weak-coupling limit, the dynamical changes induced by the coupling
occur on long time scales compared to the period of the intrinsic oscillations 
and one can thereby invoke averaging techniques.
With reference to the model Eq.~(\ref{eq:LIF4n}), it is convenient
to expand the coupling term into Fourier modes,
\begin{equation}
\label{eq:coup1}
  \Gamma(\phi_i) S(\phi_j) = \sum_{n,m} \tilde \Gamma_n \tilde S_m
  \mathrm{e}^{-2\pi \ii (n \phi_i + m\phi_j)}  \; .
\end{equation}
By assuming that only 1:1 resonances matter and retaining the secular terms.
i.e. those for which $m=-n$, one obtains
\begin{equation}
\label{eq:coup2}
  \Gamma(\phi_i) S(\phi_j) \simeq \sum_{n} \tilde \Gamma_n \tilde S_{-n}
  \mathrm{e}^{-2\pi \ii n(\phi_i - \phi_j)} = G(\phi_i-\phi_j)\;,
\end{equation}
so that Eq.~(\ref{eq:conv}) is obtained since
$\tilde S_{-n}=\tilde S_n$.

\section{Derivation of the coupling function of model C}
\label{app:KDMcf}

Using Eq.~(\ref{eq:coup3}) together with Eqs.~(\ref{eq:conv},\ref{eq:PRC}) 
the convolution integral can be written as
\begin{align}
G(\xi)&=\frac{\alpha^2}{\tilde a} \int \mathrm{e}^{\frac{\psi+\xi}{\nu}}
\mathrm{e}^{-\frac{\alpha\psi}{\nu}}(A\psi+B)\dd\psi \nonumber\\[2ex]
&-\frac{\nu^2}{\tilde{a}}\int \mathrm{e}^{\frac{\psi+\xi}{\nu}}\dd\psi
=\frac{\alpha^2}{\tilde a} I_1-\frac{\nu^2}{\tilde{a}} I_2
\nonumber\;,
\end{align}
where  $A=(1-\mathrm{e}^{-\alpha\tau})^{-1}$, 
$B=\mathrm{e}^{-\alpha\tau}(1-\mathrm{e}^{-\alpha\tau})^{-2}$, and $\tilde a = a + g\nu$.
Taking into account that $\psi+\xi$ shall be understood as taken modulo one, 
we write 
\begin{align}
I_1&=\mathrm{e}^{\xi\tau}\int_0^{1-\xi}\mathrm{e}^{-(\alpha-1)\psi\tau}(A\psi+B)\dd\psi
\nonumber\\[2ex]
&+\mathrm{e}^{\xi\tau}\mathrm{e}^{(\alpha-1)\tau}
\int_{1-\xi}^1 \mathrm{e}^{-(\alpha-1)\psi\tau}(A\psi+B]\dd\psi\;,\nonumber
\end{align}
\begin{align}
I_2&=\mathrm{e}^{\xi\tau}\int_0^{1-\eta}\mathrm{e}^{\psi\tau}\dd\psi
+\mathrm{e}^{\xi\tau}\mathrm{e}^{-\tau}
\int_{1-\xi}^1 \mathrm{e}^{\psi\tau}\dd\psi
\nonumber\;.
\end{align}
The further integration is straightforward;  it yields Eq.~(\ref{eq:Gphi}),
where the coefficients are given by the following expressions,
\begin{equation}
\begin{array}{lcl}
g_1 = -\displaystyle\frac{\nu\alpha^2 (\mathrm{e}^\tau-1)}
{(a+g\nu)(\alpha-1)(\mathrm{e}^{\alpha\tau}-1)}\;,\\[3ex]
g_2 = \displaystyle\frac{1}{1-\mathrm{e}^{-\alpha\tau}} +
\displaystyle\frac{\nu}{\alpha-1} \;, \\[3ex]
g_3 = \displaystyle\frac{\nu^2\alpha^2}{(a+g\nu)(\alpha-1)^2}\;, \quad
g_4 = \displaystyle\frac{\nu^3 (\mathrm{e}^\tau -1)}{a+g\nu}\;.
\end{array}
\label{eq:gcoeffs}
\end{equation}

\section{Linear stability of the splay state}
\label{sec:linsplay}

\subsection{Model A}
\label{sec:app1}
The weak-coupling limit of the splay state in this setup has been first studied in
\cite{Abbott-vanVreeswijk-93} and more recently extended to a broader class of pulse-coupled
integrate-and-fire systems in \cite{Olmi-Politi-Torcini-10}.
We start from Eq.~(\ref{eq:cont}) with the flux
\begin{equation}
J(\phi,t) = \left[ \nu + g \mathrm{e}^{\phi/\nu}\varepsilon(t)\right] P(\phi,t) 
\label{flux}
\end{equation}
and with the boundary condition $J(0,t)=J(1,t)$. The evolution
equation for the field is
\begin{equation}
\label{eq:Eth}
  \ddot \varepsilon(t) +2\alpha\dot \varepsilon(t)+\alpha^2 \varepsilon(t)=  
\alpha^2 (J(1,t)-E_0) \, .
\end{equation}
The splay state corresponds to $P_0 = 1$, $\varepsilon=0$, and $J_0=E_0=\nu$.

Upon introducing the perturbation $j(\phi,t)$ to the steady flux $J_0=\nu$,
i.e. writing $J(\phi,t) =  \nu + j(\phi,t)$,
the evolution equations (\ref{eq:cont},\ref{flux},\ref{eq:Eth})
can be linearised, yielding
\begin{eqnarray}
\frac{\partial j}{\partial t} = 
\frac{\nu g}{\tilde a} \mathrm{e}^{\phi /\nu} \frac{\dd \varepsilon}{\dd t} 
- \nu \frac{\partial j}{\partial \phi} \;, \\[1ex] 
  \ddot \varepsilon(t) +2\alpha \dot \varepsilon(t) +
\alpha^2 \varepsilon(t)= \alpha^2 j(1,t) \;.
\end{eqnarray}
Using the standard Ansatz $j(\phi,t) = j_f(\phi) \exp(\mu t)$ and 
$\varepsilon(t) = \varepsilon_f \exp(\mu t)$, and imposing the boundary 
condition $j_f(0)=j_f(1)$, 
one obtains the eigenvalue equation
\begin{equation}
\left({\rm e}^{\mu/\nu}-1\right) (\mu + \alpha)^2 = \frac{g \alpha^2 \mu}{\tilde a}
 \int_0^1 \dd\phi \mathrm{e}^{(1+\mu)\phi/\nu} \;.
\label{charac}
\end{equation}

We now investigate the weak coupling limit $g\ll 1$.
For $g=0$, two eigenvalues are obtained by solving 
$(\mu + \alpha)^2 = 0$,
i.e. $\mu=-\alpha$ is a double degenerate solution.
Besides, the spectrum consists of an infinite set 
of purely imaginary eigenvalues,
$\mu = 2 \pi \ii n \nu$, $n \ne 0$,
which are most important for determination of stability.
In the small $g$ limit one can assume 
$ \mu_n = 2 \pi \ii n \nu + g \delta_n $.
Upon replacing in Eq.~(\ref{charac}), we obtain
\begin{equation*}
 \delta_n (2\pi \ii n\nu + \alpha)^2 = \frac{2\pi \ii n \alpha^2\nu^2 }{\tilde a}
 \int_0^1 \dd\phi \mathrm{e}^{(1/\nu+2\pi \ii n)\phi}\;.
\label{charac3}
\end{equation*}
Computing the integral, one obtains the final Eq.~(\ref{eq:eigenvdel2}).

\subsection{Model B}
\label{sec:app2}
Here, we refer to model (\ref{eq:LIF4n}). In the thermodynamic limit, 
the sum over all oscillators transforms into an integral, and the 
expression for the probability flux takes the form
\begin{equation}
J(\phi,t) = 
\left[ \nu + g \Gamma(\phi) S_P(t) \right] P(\phi,t) \;,
\label{eq:contn}
\end{equation}
where
\begin{equation}
S_P(t)  = \int_0^1 \dd\psi S(\psi)P(\psi,t)\;  ,
\label{eq:int0}
\end{equation}
while the boundary condition reads 
\begin{eqnarray}
[\nu+g\Gamma(1) S_P(t)]P(1,t) = [\nu+g\Gamma(0) S_P]P(0,t) \, .
\label{eq:boundw}
\end{eqnarray}

At variance with the previous case, the stability can be assessed by  
just linearizing the above equation, without the need of including the field dynamics.
The problem can be formally solved for arbitrary coupling strength

Starting from Eqs.~(\ref{eq:contn},\ref{eq:int0})  with the boundary condition
(\ref{eq:boundw}), we set $P(\phi,t) = 1 + p(\phi,t)$, where $p(\phi,t)$ represents
a perturbation around the homogeneous solution.
The linearized equation writes
\begin{equation}
\frac{\partial p}{\partial t} = -\nu \frac{\partial p}{\partial \phi} 
-g \Gamma' (\phi) S_p\;,
\label{eq:lincont0}
\end{equation}
where the prime denotes derivation with respect to $\phi$ and 
$S_p$ is defined analogously to $S_P$, see Eq.~(\ref{eq:int0}); 
notice also that $S_P=0$ in the splay state. 
The boundary condition becomes
$$ \nu[p(1,t)-p(0,t)] =-g S_p \Delta \Gamma\;,$$ 
where $\Delta \Gamma = \Gamma(1)-\Gamma(0)$.

Next, we introduce the usual Ansatz, 
$p(\phi,t) = \rho(\phi) {\rm e}^{\mu t}$, obtaining 
\begin{equation}
 \nu \frac{\dd \rho}{\dd \phi} = -\mu \rho -g \Gamma' (\phi) S_\rho \;,
\label{eq:lincont2}
\end{equation}
where $S_\rho$ is defined analogously to $S_P$, see Eq.~(\ref{eq:int0}).
By assuming that $\rho(\phi) = \rho_0(\phi) \exp(-\mu \phi/\nu)$,
we find that
\begin{equation}
\rho_0(\phi) = -\frac{g}{\nu} S_\rho I_\mu(\phi) + C \;,
\label{eq:lincont3}
\end{equation}
where 
\begin{equation}
I_\mu(\phi) = \int_0^\phi \dd \xi \Gamma'(\xi){\rm e}^{\mu \xi/\nu} \;.
\label{eq:defint}
\end{equation}
The integration constant can be determined from the boundary condition
\begin{equation}
C  = \frac{g}{\nu} \frac{{\rm e}^{-\mu/\nu} I_\mu(1)  - \Delta \Gamma}
{{\rm e}^{-\mu /\nu} -1} S_\rho \;.
\label{eq:bc3}
\end{equation}
As a result,
\begin{equation}
\rho(\phi) = \frac{g}{\nu} {\rm e}^{-\mu \phi/\nu}
\left[ \frac{ {\rm e}^{-\mu/\nu} I_\mu(1) -  \Delta \Gamma} 
{{\rm e}^{-\mu /\nu}-1} - I_\mu(\phi)  \right] S_{\rho} \;.
\label{eq:lincont5}
\end{equation}
The eigenvalue equation is finally obtained by multiplying $\rho(\phi)$ by $S(\phi)$
and integrating over $\phi$ to obtain $S_\rho$:
\begin{equation}
g \frac{{\rm e}^{-\mu/\nu} I_\mu(1) -\Delta \Gamma}{{\rm e}^{-\mu/\nu}-1}
\langle {\rm e}^{-\mu\phi/\nu}\rangle_S -
g\langle I_\mu(\phi){\rm e}^{-\mu\phi/\nu}\rangle_S = \nu\;,
\label{eq:eigen1}
\end{equation}
where $\langle \cdot \rangle_S$ denotes the integral over the dummy variable $\phi$ after having
been multiplied by $S(\phi)$.

In the  weak coupling limit, the second addendum in the l.h.s. of the above equation
can be neglected, while the first one can be properly handled by assuming 
$\mu_n = 2 \pi \ii n \nu  + g \delta_n$ in the numerator 
(and $ \mu_n = 2 \pi \ii n \nu$ everywhere else). 
As a result, the eigenvalue equation simplifies to
\begin{equation}
 \delta_n =  - \left[ \tilde \Gamma'_n -\Delta \Gamma \right]  \tilde S_{n}^* \;,
\end{equation}
since Eq.~(\ref{eq:defint}) reduces to the Fourier transform of $\Gamma'$, while
$\langle \mathrm{e}^{-\mu\phi/\nu} \rangle_S$ reduces to the conjugate of the transform
of $S$.
From Eq.~(\ref{eq:PRC}), it follows that
\begin{equation}
  \tilde \Gamma_n = \int_0^1 \dd\phi\, \Gamma(\phi) \mathrm{e}^{2\pi \ii n \phi}
  = \frac{\nu}{\tilde a} \frac{\mathrm{e}^{1/\nu} -1}{1/\nu+2\pi \ii n}\; ,
\label{eq:Gnp}
\end{equation}
and, accordingly,
\begin{equation}
  \tilde \Gamma'_n = \frac{1}{\tilde a} \frac{\mathrm{e}^{1/\nu} -1}{1/\nu+2\pi \ii n} \;,
\label{eq:Gnpp}
\end{equation}
so that
\begin{equation}
  \tilde \Gamma'_n -\Delta \Gamma = 
-\frac{\mathrm{e}^{1/\nu} -1}{\tilde a} \frac{2\pi \ii n\nu} {1/\nu+2\pi \ii n} \, .
\label{eq:DGnp}
\end{equation}
By further noticing that
\begin{equation}
  \tilde S_n = \int_0^1 \dd\phi\, S(\phi) \mathrm{e}^{2\pi \ii n \phi} 
  = \frac{\alpha^2\nu}{(\alpha-2\pi \ii n\nu)^2}\;.
\label{sn}
\end{equation}
we finally obtain Eq.~(\ref{eq:eigenvdel2}).

\subsection{Model C}
\label{sec:app3}
In the thermodynamic limit, Eq.~(\ref{eq:daido}) can be written as
\begin{eqnarray}
\label{eq:daido2}
  \dot{\phi} &=& \nu + g\! \int \dd\psi G(\phi -\psi)P(\psi)  \;,
\end{eqnarray}
or, using the Fourier representation, as
\begin{equation}
\label{eq:daido3}
  \dot{\phi}  = \nu + g \sum_n \tilde G_n \tilde P_{n} \mathrm{e}^{-2\pi \ii n \phi} \;.
\end{equation}
Accordingly, the continuity equation becomes
\begin{equation}
\frac{\partial P}{\partial t} = -\frac{\partial }{\partial \phi} 
\left[ \left( \nu + g \sum_n \tilde G_n \tilde P_{n} \mathrm{e}^{-2\pi \ii n \phi} \right )
P(\phi,t) \right] \;.
\label{eq:contn2}
\end{equation}
We now linearize Eq.~(\ref{eq:contn2}) around the splay solution $P_0(\phi)=1$, by
assuming $P(\phi,t) = 1 +p(\phi,t)$.  Since the mode amplitudes of the equilibrium
solution for $n \ne 0$ are all equal to zero,  
\begin{equation}
\frac{\partial p}{\partial t} = -\nu \frac{\partial p}{\partial \phi}
+ 2\pi \ii g \sum_{n\ne 0} n \tilde G_n 
\tilde p_{n} \mathrm{e}^{-2\pi \ii n \phi} \;.
\label{eq:contn3}
\end{equation}
At variance with the previous setups, one can easily solve the continuity equation
by just going in Fourier space, as this change of variables diagonalizes
the evolution equation for any parameter value
\begin{equation}
\frac{\dd{\tilde p}_{n}}{\dd t} = 2\pi \ii n\left[ \nu 
+ g \tilde G_{n} \right] \tilde p_{n}\;.
\label{eq:contn4}
\end{equation}
By recalling that $\mu_n \equiv 2 \pi \ii n \nu  + g \delta_n$,
\begin{equation}
\delta_n = 2\pi \ii n \tilde G_{n} = 
2\pi \ii n \tilde \Gamma_{n} \tilde S_n^* \;,
\label{eq:contn5}
\end{equation}
which, in the case of the LIF model, coincides with
Eq.~(\ref{eq:eigenvdel2}).

\section{Linear stability of the synchronous state}
\label{sec:linsyn}

\subsection{Model A}
From Eq. (52) in \cite{Olmi-Politi-Torcini-10},
the period $T$ for the an ensemble of LIF oscillators with $\alpha$-pulses
is determined by the implicit condition
\begin{equation}
a(1\!-\!\mathrm{e}^{-T})\! +\! g\! \left[
\frac{\mathrm{e}^{-T}\!-\!\mathrm{e}^{-\alpha T}}{\alpha-1}
(V+Q) \!-\!T\mathrm{e}^{-\alpha T}Q
\right]\!=\!1\;,
\label{eq:LIFperiod}
\end{equation}
where 
\begin{equation}
Q = \frac{\alpha^2/(\alpha-1)}{1-\mathrm{e}^{-\alpha T}}\quad ,\quad
V = \frac{\alpha^2 T \mathrm{e}^{-\alpha T}}{(1-\mathrm{e}^{-\alpha T})^2} \, .
\end{equation}
For $g=0$, the period is equal to $\tau=-\ln(1-1/a)\equiv 1/\nu$, cf. Eq.~(\ref{eq:nu2});
let us denote with $Q_0$, $V_0$ the corresponding values of $Q$ and $V$.
In the small $g$ limit, we can assume $T=\tau+\delta T$, where $\delta T$ is small, and
expand the first term in Eq.~(\ref{eq:LIFperiod}) (the second term is already of order $g$), 
obtaining
\begin{equation}
\delta T =\! -\frac{g}{a}\left[
\frac{1\!-\!\mathrm{e}^{-(\alpha-1)\tau}}{\alpha-1}(V_0+Q_0)
\!-\!\tau\mathrm{e}^{-(\alpha-1)\tau }Q_0\right ]\;.
\end{equation}
By replacing the expressions for $V_0$ and $Q_0$, we finally obtain
Eqs.~(\ref{eq:LIFtau},\ref{eq:B}.

The stability of the limit cycle is determined by the exponent~\cite{Olmi-Politi-Torcini-10}, 
\begin{equation}
\label{eq:LIFsy}
\lambda = \frac{1}{T} \ln \frac{a+gV}{a-1+gV} -1\;.
\end{equation}
By now expanding for $g\ll 1$, we obtain, up to the first order,
\begin{equation}
\label{eq:LIFsy2}
\lambda = -\frac{\delta T}{\tau} - \frac{gV_0}{\tau a(a-1)}\;.
\end{equation}
With the help of Eq.~(\ref{eq:LIFtau}) and recalling that 
$\mathrm{e}^\tau-1=1/(a-1)$, one obtains the final expression for
the Lyapunov exponent that is reported in Eq.~(\ref{eq:LIFsy3}).

\subsection{Model B}

Here, we determine the period and determine the stability of the fully 
synchronous solution of the model (\ref{eq:LIF4n}).
In the weak coupling limit, the period can be estimated through 
a perturbative calculation, by setting $\phi = \nu t + \beta(t)$ in 
Eq.~(\ref{eq:LIF4n}) and retaining the leading order,
\begin{equation}
\label{eq:winb}
  \dot{\beta} =  g \Gamma(\nu t) S(\nu t) \, .
\end{equation}
The period $T$ can be then obtained by solving the above equation and imposing
\begin{equation}
\label{eq:newper}
   \nu T + \beta(\tau) = 1\;,
\end{equation}
so that
\begin{equation}
\label{eq:newperi2}
   \delta T = - \tau \beta(\tau) \, .
\end{equation}
$\beta(\tau)$ can be determined by integrating Eq.~(\ref{eq:winb}) that
can be written as,
\begin{equation}
\label{eq:winrbeta}
  \dot{\beta} =  \frac{g \alpha^2}{a}  \left [
\frac{\nu t\mathrm{e}^{-(\alpha-1) t}}{1-\mathrm{e}^{-\alpha/\nu}} + 
\frac{\mathrm{e}^{-\alpha/\nu} \mathrm{e}^{-(\alpha-1)t}}{(1-\mathrm{e}^{-\alpha/\nu})^2} 
- \frac{\nu^2}{\alpha^2} \mathrm{e}^t 
\right ] \,,
\end{equation}
where we have used that $\tilde a = a$ in the weak coupling limit.
By replacing the integral of this equation into Eq.~(\ref{eq:newperi2}),
one obtains the expression reported in Eq.~(\ref{eq:wintau}).

As for the stability, the tangent space evolution writes
\begin{equation}
\label{eq:wint}
  \frac{\dd\delta\phi_i}{\dd t} =  g \Gamma'(\phi_i) \langle S\rangle \delta \phi_i
  +  g \Gamma(\phi_i) \frac{1}{N} \sum S'(\phi_j) \delta \phi_j \;.
\end{equation}
If all the oscillators are synchronized, we can drop the index dependence in the phase
space dynamics,
\begin{equation}
\label{eq:wint2}
 \frac{\dd\delta\phi_i}{\dd t}  =  g \Gamma'(\phi) S(\phi) \delta \phi_i
  +  g \Gamma(\phi) \frac{S'(\phi)}{N} \sum_j \delta \phi_j \;.
\end{equation}
The stability can be assessed by introducing the 
variables $\theta_i = \delta \phi_i - \delta \phi_1$ with
$i\ge 2$ \cite{Strogatz-94} 
(the sum of all $\delta \phi_i$ gives a missing equation which is known
to yield the zero exponent and we thereby avoid considering it),
\begin{equation}
\label{eq:wint3}
  \dot\theta_i =  g \Gamma'(\phi) S(\phi) \theta_i \;.
\end{equation}
Since $\Gamma(\phi)$ is discontinuous for $\phi=1$, its derivative has a delta contribution
that has to be properly included in the computation of the Floquet exponent. The final result is
\begin{equation}
\label{eq:wint5}
\lambda =  \frac{1}{T} \ln \frac{\theta(T)}{\theta(0)} +
 \frac{1}{T} \ln \frac{\nu+g\Gamma(0)S(0)}{\nu+g\Gamma(1)S(1)} \;.
\end{equation}
In the limit $g\ll 1$, taking into account that $S(0)=S(1)$  the above equation reduces to
\begin{equation}
\label{eq:wint6}
\lambda =  \frac{1}{\tau} \ln \frac{\theta(\tau)}{\theta(0)} + G[\Gamma(0)-\Gamma(1)]S(0) \; .
\end{equation}
One can then determine $\theta(\tau)$ by integrating Eq.~(\ref{eq:wint3}) with the
same philosophy as for Eq.~(\ref{eq:winb}). 
As a result the same expression as (\ref{eq:LIFsy3}) is obtained for $\lambda$.

\subsection{Model C}

The determination of the period is pretty straightforward: it can
be obtained by setting the
argument of the interaction function $G$ equal to zero
\begin{equation}
\label{eq:daiper0}
   \frac{1}{T} =  \nu + g G(0) \;,
\end{equation}
so that, for the LIF oscillators,
\begin{equation}
\label{eq:daiper}
   \delta T =  - g G(0)\tau^2 = g(g_1g_2+g_3-g_4)\tau^2 \;.
\end{equation}
Upon replacing the expressions for $g_1$, $g_2$, $g_3$ and $g_4$ reported in
appendix~\ref{app:KDMcf}, one can verify that the above equation coincides with Eq.~(\ref{eq:wintau}).

Next, we linearize the equations of motion, obtaining
\begin{equation}
\label{eq:daistab}
  \frac{\dd\delta\phi_i}{\dd t}=  
  g G'(0)\delta \phi_i- \frac{g}{N} G'(0) \sum_j -\delta \phi_j \;.
\end{equation}
The stability can be determined by again introducing the variables
$\theta_i=\delta \phi_i-\delta \phi_1$, which satisfy the following equation
\begin{equation}
\label{eq:daistab2}
  \dot\theta_i=  g G'(0)\theta_i \;,
\end{equation}
so that the stability is controlled by the sign of $G'(0)$.

\section{Computation of the first Fourier mode of the probability 
distribution in SCPS state}
\label{sec:app4}

Substituting Eq.~(\ref{eq:fixfou7}) into Eq.~(\ref{eq:fixfou8}), 
we obtain 
\begin{equation}
\tilde P_1^2 = \frac{\hat\W_0-g\tilde G_2}{\hat\W_0+g\tilde G_2} \cdot
\frac{\delta \Omega -g\tilde G'_1 \delta \mu}{\hat\W_0} \;.
\label{eq:fixfou9}
\end{equation}
Condition $\mbox{Im}(\tilde P_1^2)=0$ yields
\begin{equation}
\delta \W = \left [(\tilde G'_1)^r  + (\tilde G'_1)^i  
\frac{g^2|\tilde G_2|^2-\hat\W_0^2}{2g\hat\W_0 \tilde G_2^i} \right] \delta \mu
=M \delta \mu \;.
\label{eq:condre2}
\end{equation}
As a result, from Eq.~(\ref{eq:fixfou9}), it follows 
that $\tilde P_1^2$ is proportional to $\delta \mu$,
\begin{equation}
\tilde P_1^2 = \frac{\hat\W_0-g\tilde G_2}{\hat\W_0+g\tilde G_2} \cdot
\frac{M -g\tilde G'_1}{\hat\W_0} \delta \mu\;.
\label{eq:p2b}
\end{equation}

To complete the computation we have to 
find $\tilde G_2$, $\tilde G_1'$ at the bifurcation point.
With the reference to the Abbot -- van Vreeswijk model, the coupling function is given by Eq.~(\ref{eq:Gphi}) 
and the role of the order parameter is played by the inverse 
pulse width $\alpha$;
the bifurcation value $\alpha_c$ is given by Eq.~(\ref{crit0}). 
Computing the first two Fourier modes of $G$, we find:
\begin{equation}
\tilde G_{1,2} = C \frac{A_{1,2}-\ii B_{1,2}} {D_{1,2}}   \, ,
\label{eq:gtilde1}
\end{equation}
where
\begin{equation*}
C = \frac{\alpha_c^2\nu^3 }{a+g\nu}(\mathrm{e}^{1/\nu} -1)
\end{equation*}
and
\begin{eqnarray}
 A_n &=& \alpha_c^2\! -\! (2\pi n\nu)^2(1\!+\!2\alpha_c) \,,\nonumber\\
 B_n &=& 2\pi n \nu \left [ \alpha_c^2\!+\!2\alpha_c\!-\!(2\pi n \nu)^2\right ]\;,\nonumber \\
 D_n &=& [\alpha_c^2 + (2\pi n\nu)^2]^2 [1+(2\pi n\nu)^2]\;.\nonumber
\label{charac3c}
\end{eqnarray}
Finally,
\begin{align*}
(\tilde G'_1)^r &= \frac{2C}{D_1}
\left (\frac{A_1}{\alpha_c}-
\frac{2\alpha_cA_1}{4\pi^2\nu^2+\alpha_c^2}+\alpha_c-4\pi^2\nu^2\right )\;,\\[2ex]
(\tilde G'_1)^i &= 
-\frac{4C\pi \nu(1+ \alpha_c)}{D_1}\;.
\label{eq:gtilde3}
\end{align*}


%

\end{document}